\documentclass[conference]{IEEEtran}
\IEEEoverridecommandlockouts
\usepackage[font=small,labelfont=bf,
   justification=justified,
   format=plain]{caption}
\usepackage{cite}
\usepackage{amsmath,amssymb,amsfonts}
\usepackage[linesnumbered,algoruled,boxed,lined]{algorithm2e}
\usepackage{amsthm}
\usepackage{graphicx}
\usepackage{textcomp}
\usepackage{xcolor}
\usepackage{amsthm}
\usepackage{graphicx}
\usepackage{dcolumn}
\usepackage{bm}
\usepackage[export]{adjustbox} 
\usepackage{enumitem}
\usepackage{booktabs}
\usepackage{makecell}
\usepackage{hyperref}
\usepackage[braket, qm]{qcircuit}
\usepackage{subcaption, wasysym}
\usepackage{xcolor}
\usepackage{todonotes}
\usepackage{soul}

\theoremstyle{definition}
\newtheorem{definition}{Definition}

\theoremstyle{definition}
\newtheorem{example}{Example}[section]
\SetKwRepeat{Do}{do}{while}%
\newcommand{\head}[1]{\noindent \textbf{#1}:}

\hypersetup{
	colorlinks,
	linkcolor={red!50!black},
	citecolor={blue!50!black},
	urlcolor={blue!80!black}
}
\def\BibTeX{{\rm B\kern-.05em{\sc i\kern-.025em b}\kern-.08em
    T\kern-.1667em\lower.7ex\hbox{E}\kern-.125emX}}
\begin{document}

\title{MUQUT: Multi-Constraint Quantum Circuit Mapping on Noisy Intermediate-Scale Quantum Computers}

\author{
    \IEEEauthorblockN{Debjyoti Bhattacharjee\IEEEauthorrefmark{1}, Abdullah Ash Saki\IEEEauthorrefmark{2}, Mahabubul Alam\IEEEauthorrefmark{2}, Anupam Chattopadhyay\IEEEauthorrefmark{1}, Swaroop Ghosh\IEEEauthorrefmark{2}}
    \IEEEauthorblockA{\IEEEauthorrefmark{1}School of Computer Science and Engineering, Nanyang Technological University, Singapore}
    \IEEEauthorblockA{\IEEEauthorrefmark{2}School of Electrical Engineering and Computer Science, Pennsylvania State University, University Park, USA}
    Email: {\emph {debjyoti001@ntu.edu.sg}}
 }

\maketitle

\begin{abstract}
Rapid advancement in the domain of quantum technologies have opened up researchers to the real possibility of experimenting with quantum circuits, and simulating small-scale quantum programs. Nevertheless, the quality of currently available qubits and environmental noise pose a challenge in smooth execution of the quantum circuits. Therefore, efficient design automation flows for mapping a given algorithm to the Noisy Intermediate Scale Quantum (NISQ) computer becomes of utmost importance. State-of-the-art quantum design automation tools are primarily focused on reducing logical depth, gate count and qubit counts with recent emphasis on topology-aware (nearest-neighbour compliance) mapping. In this work, we extend the technology mapping flows to simultaneously consider the topology and gate fidelity constraints while keeping logical depth and gate count as optimization objectives. We provide a comprehensive problem formulation and multi-tier approach towards solving it. The proposed automation flow is compatible with commercial quantum computers, such as IBM QX and Rigetti. Our simulation results over 10 quantum circuit benchmarks, show that the fidelity of the circuit can be improved up to 3.37$\times$ with an average improvement of 1.87$\times$.
\end{abstract}

\section{Introduction}
\noindent Quantum computation~\cite{NC:2000} promises to expand the reach of computing beyond classical - both theoretically and practically. It is conjectured that there are problems  belonging to Bounded-Error Quantum Polynomial (BQP) class, which cannot be solved efficiently in a classical computer. There are already several problems, most notably, Shor's number factorization and Boson Sampling that demonstrate superpolynomial speed-up over the best known classical algorithms. Consequently, a massive research effort is underway to develop practical, scalable quantum computers.
To enable convenient programming for quantum computers, Microsoft has open-sourced Q\#~\cite{Qsharp} and IBM has provided selective remote access to small-scale quantum computers~\cite{Qiskit}. These frameworks present opportunity for researchers to describe a quantum algorithm and test its outcome by running on a practical system.
However, the scalability of quantum computers is hindered due to their extreme susceptibility to decoherence and noise. According to Threshold theorem, Quantum Error-Correcting Codes (QECC) can only provide robustness up to a level of local noise (i.e., threshold) \cite{fowler_surface12}. The resource requirement for QECC may scale up faster than the number of computing qubits, eventually posing as a roadblock. 

In the interim, there remains an interesting possibility to explore Noisy Intermediate-Scale Quantum (NISQ) computers~\cite{preskillnisq} to solve `useful' computations demonstrating the efficacy of scalable quantum computers, even though at a small scale. Needless to mention, the NISQ era also presents an opportunity for the software tool chain to mature and prepare well ahead of the arrival of large-scale quantum computing. A typical tool chain consists of the programming language to describe quantum circuits and algorithms,  synthesis and technology mapping phases. We focus on the challenge of technology mapping. 

While mapping a given quantum algorithm on NISQ computer, primarily two sets of constraints are considered - topology and fidelity. On one hand, topology-aware constraints in state-of-the-art literature considered 2D topology (barring a few exceptions). On the other hand, fidelity-aware mapping flows began with a quantum circuit that is topology-compliant. To the best of our knowledge, the interplay between these two constraints are not explored. In this paper, we discuss all variants of topology and fidelity constraints and present MUQUT - Multi-Constraint Quantum Circuit Mapping flow. We have made following novel contributions.
\begin{itemize}
    \item A multi-constraint quantum circuit mapping problem formulation for NISQ-era quantum computers. 
    \item A multi-tier technology mapping flow, combining optimal and heuristic solutions for various sub-problems, is presented (section~\ref{sec:method}). 
    \item Demonstrative examples and benchmarking based on the commercial quantum architectures (IBM, Rigetti) to validate our advances (section~\ref{sec:res}).
\end{itemize}

\section{Background and Motivation}\label{sec:background}
In quantum computing, the operations take place on \emph{qubits}, which are a linear combination of the conventional Boolean states in the two dimensional complex Hilbert space. Each operation on these qubits can be defined by a unitary matrix~\cite{NC:2000} represented by means of quantum gates. 
\begin{definition}(Quantum gate) A {quantum gate} over the inputs \mbox{$X=\{x_1,\dots , x_n\}$} consists of a single target line~$t\in X$ and, one or more control line(s)~$c\in X$ with~\mbox{$t\neq c$}.
\end{definition}
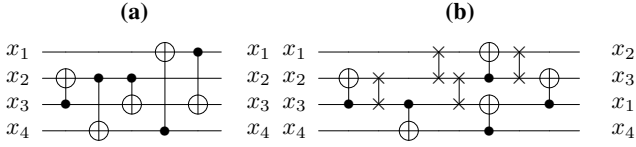
\begin{figure}[t]
	\begin{subfigure}[t]{0.5\columnwidth}
		\centering 
		\caption{} 
		\label{fig:circorig}
		\scalebox{0.9}{
			\Qcircuit  @C= 2mm @R=1mm {
				\lstick{x_1} & \qw & \qw & \qw & \targ & \ctrl{2} & \qw & \rstick{x_1}  \\
				\lstick{x_2} & \targ & \ctrl{2} & \ctrl{1} & \qw & \qw & \qw & \rstick{x_2}  \\
				\lstick{x_3} & \ctrl{-1} & \qw & \targ & \qw & \targ & \qw & \rstick{x_3}  \\
				\lstick{x_4} & \qw & \targ & \qw & \ctrl{-3} & \qw & \qw & \rstick{x_4}  \\
			}
		}
	\end{subfigure}
	\begin{subfigure}[t]{0.45\columnwidth}
		\centering
		\caption{} 
		\label{fig:circnn}
		\scalebox{0.9}{
			\Qcircuit  @C= 3mm @R=1mm {
				\lstick{x_1} & \qw & \qw & \qw & \qswap & \qw & \targ & \qswap & \qw & \qw & \rstick{x_2} & \\
				\lstick{x_2} & \targ & \qswap & \qw & \qswap \qwx & \qswap & \ctrl{-1} & \qswap \qwx & \targ & \qw & \rstick{x_3} & \\
				\lstick{x_3} & \ctrl{-1} & \qswap \qwx & \ctrl{1} & \qw & \qswap \qwx & \targ & \qw & \ctrl{-1} & \qw & \rstick{x_1} & \\
				\lstick{x_4} & \qw & \qw & \targ & \qw & \qw & \ctrl{-1} & \qw & \qw & \qw & \rstick{x_4} & \\
			}  
		}
	\end{subfigure}
	\caption{(\subref{fig:circorig})~A quantum circuit with five CNOT gates and four levels. (\subref{fig:circnn})~The nearest neighbour
	compliant circuit for LNN topology.} 
	\label{fig:example_naive_qua}
\end{figure}
\begin{definition}[Quantum circuit]
	A quantum circuit, defined over $n$-qubits $q_1$, $q_2$,...,$q_n$ is a series of levels $L_i$, where each level $L_i$ consists of a set of quantum gates $G_i^1$, $G_i^2$, $\cdots$, $G_i^k$ with each gate $G_i^j$ operating on one or more qubits.
\end{definition}
Any two pair of gates $G_i^j$ and $G_i^k$ in a level $L_i$ do not operate on any common qubit and therefore can be executed in parallel. At a logical level, we consider that each level $L_i$ takes one cycle to execute. A quantum circuit with $k$~levels has a delay of $k$~cycles.
\begin{example}
	Fig.~\ref{fig:circorig} shows a quantum circuit with 5 CNOT gates. The circuit has 4 levels, $L_1 = \{G_1\}$, $L_2 = \{G_2\}$, $L_3 = \{G_3,G_4\}$ and $L_4 = \{G_5\}$.
\end{example}
\begin{definition}[Clifford group] The Clifford group is a set of special kind of quantum gates (G) which satisfies 
	\begin{equation}
	G^{\dag}PG \in P^{(\oplus)n}
	\end{equation}
	where $P$ represents Pauli-gate and  $P \in \{I, X, Y, Z\}$. \dag represents the self inverse of a gate.  The Clifford group is composed with \textit{H} gate, Pauli's matrices $\{X, Y, Z \}$ and \textit{S} gate along with \textit{CNOT} gate.
\end{definition}
Each logical gate in Clifford group~(NOT, CNOT, H, Z, X, Y and S) is transversal. Transversal operators do not propagate errors between the lines within the same encoded block of QECC. Any quantum circuit built over only transversal gates are inherently fault tolerant~\cite{Gottesman}. 

\subsection{Nearest Neighbor Compliance}\label{ssec:nn_compliance}
A major challenge towards the realization of practical and scalable quantum computing is to achieve quantum error correction. Long-distance interacting qubits are particularly susceptible to the noise. Therefore, prominent quantum technologies and quantum error correction codes, e.g., surface codes~\cite{fowler_surface12} require that the quantum gates must be formed with a nearest neighbor interaction. In the resulting circuits, the interacting qubits may form a chain, as in a 1D qubit layout, and therefore, these circuits are referred to as Linear Nearest Neighbor (LNN) circuits. 

Given a quantum gate with $m$-control lines $l_1, ... , l_m$ and target line $l_t$, qubits $q_l$ and $q_{t}$ have to be  nearest-neighbors, $ 1 \le l \le m$. For level $L_i$,
we define \textbf{interaction} $I_i$ as the set of nearest neighbors for the all the gates in $L_i$.  
\begin{example}
	For $L_3$ of Fig.~\ref{fig:circorig}, the interaction $I_3$ is \mbox{$I_3 = \{x_2-x_3, x_1-x_4 \}$}. 
\end{example}
\begin{figure}[t]
    \begin{subfigure}[t]{0.5\columnwidth}
	\centering
	\caption{}
	\label{fig:ibmq16}
	\includegraphics[width=1.7in]{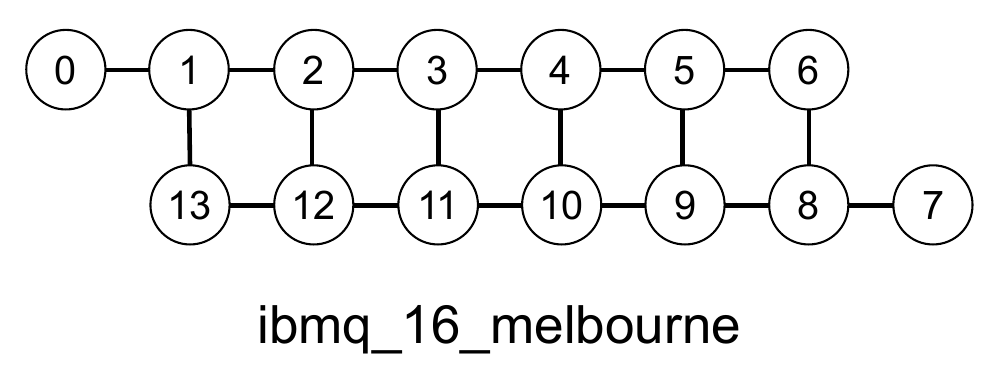}
	\end{subfigure}
	\begin{subfigure}[t]{0.49\columnwidth}
	\centering
	\caption{}
	\label{fig:rig}
	\includegraphics[width=0.80\textwidth]{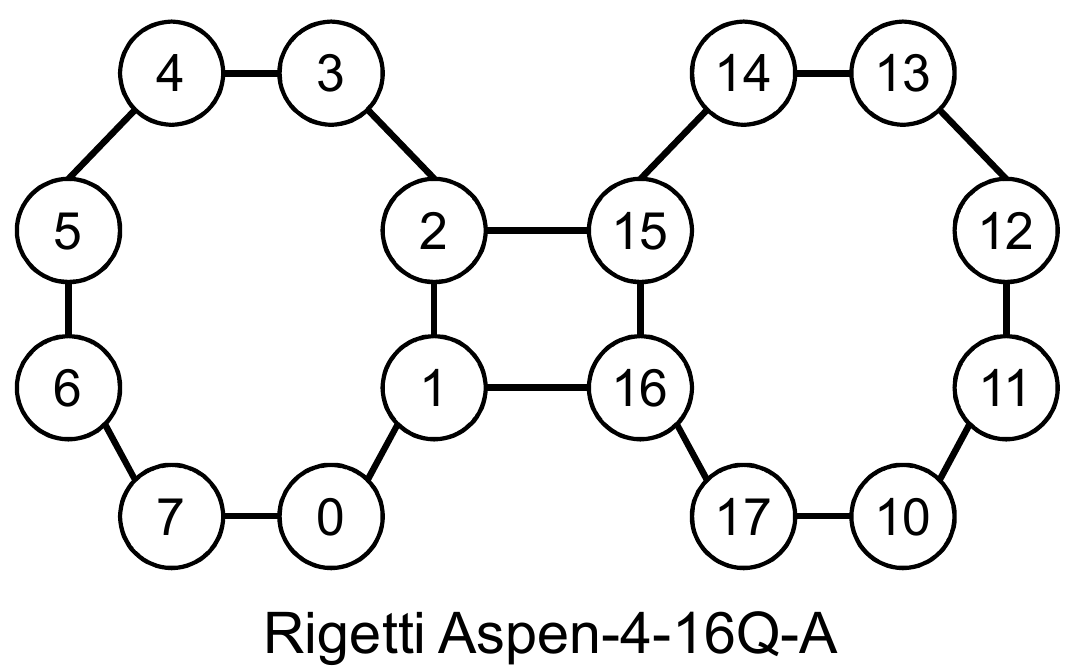}
	\end{subfigure}

	\begin{subfigure}[t]{\columnwidth}
	\caption{}
    \label{tab:error}
    \centering
{  \scriptsize

\setlength{\tabcolsep}{2pt}
\begin{tabular}{rccc||rcrc}
\hline
   \multicolumn{1}{r}{\textbf{Qubit}}  & \textbf{GE} &  \textbf{T1} & \textbf{T2} & {\textbf{Pair}} & \textbf{MGE} & \textbf{Pair} & \textbf{MGE}\\
\hline
Q0  & 3.55     & 77.30 & 22.13  & Q1-Q0 & 0.04 & Q11-Q12 & 0.07\\
Q1  & 12.8 & 65.98 & 81.72  & Q1-Q2 & 0.04 & Q12-Q2 & 0.1 \\
Q2  & 12.0 & 55.00 & 109.26 & Q2-Q3 & 0.05 & Q13-Q1 & 0.18\\
Q3  & 2.42 & 77.70 & 68.80  & Q4-Q3 & 0.04 & Q13-Q12 & 0.09\\
Q4  & 4.31 & 60.34 & 23.11  & Q4-Q10 & 0.04 \\
Q5  & 8.15 & 13.57 & 24.13  & Q5-Q4 & 0.05 \\
Q6  & 3.65 & 60.78 & 66.15  & Q5-Q6 & 0.06 \\
Q7  & 3.96 & 41.61 & 75.70  & Q5-Q9 & 0.11 \\ 
Q8  & 4.79 & 45.34 & 63.07  & Q6-Q8 & 0.05 \\
Q9  & 30.0 & 34.68 & 24.73  & Q7-Q8 & 0.04 \\
Q10 & 4.29 & 52.26 & 87.98  & Q9-Q8 & 0.32 \\
Q11 & 4.46 & 57.56 & 101.94 & Q9-Q10 & 0.31 \\
Q12 & 8.08 & 57.81 & 105.39 & Q11-Q3 & 0.05 \\
Q13 & 13.5 & 19.96 & 28.08 & Q11-Q10 &  0.07 \\ \hline
\end{tabular}
}
	\end{subfigure}
	\caption{(\subref{fig:ibmq16})~Topology graph and (\subref{tab:error})~Error specification of IBMQ\_16\_Melbourne. GE and MGE stand for single-qubit and multi-qubit gate error respectively. GE scale is $\times 10^{-3}$, and T1 and T2 times are in $\mu s$. Error data has been taken on 13-July-2019. (\subref{fig:rig}) Topology graph of Rigetti Aspen.}
\end{figure}
Conversion of a quantum circuit to LNN can be achieved by inserting SWAP gates that make all control lines and target lines adjacent. More precisely, a cascade of adjacent SWAP gates can be inserted in front of each gate $g$ with non-adjacent circuit lines in order to shift the control line of $g$ towards the target line, or vice versa, until they are adjacent. This is shown using the following example.

\begin{example}\label{example:naive}
	Consider the circuit depicted in Fig.~\ref{fig:circorig} consisting of gates~$g_1$, $g_2$, $g_3$, $g_4$ and $g_5$, numbered from left to right. As can be seen, gates~$g_2$, $g_4$, and $g_5$ are non-adjacent. Thus, in order to make this circuit nearest neighbour compliant, SWAP gates are inserted, as shown in Fig.~\ref{fig:circnn}.
\end{example}

Quite a few works have been done in recent past to convert a quantum circuit to LNN by introducing additional gates, which, naturally worsens the circuit performance by increasing the logical depth and gate count. To balance that effect, heuristic approaches~\cite{wille_aspdac16,mazdar_nnmmd16,Alireza2014,amlan_graph_partitioning,maslov_placement,bhattacharjee2017depth} and exact algorithms~\cite{wille_exact_nn} are proposed. It is pointed out ~\cite{amlan_graph_partitioning} that the problem of NN-compliant quantum circuit construction is NP-complete. Hence, it is unlikely that this problem can be solved optimally for large instances.
To the best of our knowledge,~\cite{maslov_placement,whitney_grid_07} were the first to look into arbitrary topologies for quantum circuits with nearest neighbor constraints. So far, most of the other works in this domain have concentrated on 1D qubit layout or 2D qubit lattice structures~\cite{wille_aspdac16,Alireza2014}.
The work presented in~\cite{whitney_grid_07} focuses on identifying the qubit topology best suited for a given quantum circuit placement. Relatively unexplored is the topic of mapping on available topologies. This particular problem has been dealt with in~\cite{maslov_placement} with examples taken from liquid state NMR molecules as the topologies. A graph partitioning-based approach (claimed to be asymptotically optimal for the case of chain nearest neighbour architecture) is proposed. Independently, efficient qubit topology identification and the mapping flows for specific interaction graphs have been done in~\cite{brierley_butterfly}. It has been proved that for cyclic butterfly network, the depth overhead for mapping a given quantum circuit to nearest neighbor is $6\log n$. Subsequently, the mapping algorithm is also derived. The commercial quantum computers, such as IBM QX and Rigetti do not operate on a linear array of qubits, as shown in Fig.~\ref{fig:ibmq16} and Fig.~\ref{fig:rig} respectively. This drives the need for developing nearest-neighbor mapping techniques that can support arbitrary topologies.

Practical setups for diverse quantum circuit topologies have been made available through~\cite{Qiskit}. Formally, the topology of qubits can be described by means of a topology graph.

\begin{definition}(Topology graph)
	A {topology graph} is an ordered pair  $T$=($T_\mathcal{V},T_\mathcal{E}$).  $T_\mathcal{V}$  is the vertex set, where each vertex $v \in T_\mathcal{V}$ represents a physical location where one qubit can reside. $T_\mathcal{E}$ is the edge-set, which contains a set of edges. An edge $e_{vw} \in T_\mathcal{E}$ indicates that qubit at location/vertex $v$ and $w$ can interact. In other words, qubits at location $v$ and $w$ are nearest-neighbors~(NN).
\end{definition}
\begin{example}
	Fig.~\ref{fig:ibmq16} shows the topology of a 14-qubit IBM QX Melbourne quantum computer~\cite{ibmqx}.  
\end{example}

\subsection{Fidelity of Quantum Computers} \label{ssec:fidelity}
Existing quantum computers are noisy and error-prone. The errors in the quantum computers can be broadly characterized in two classes: (a) gate errors and (b) decoherence. Due to the gate errors, the final state of the operation deviates from the ideal state. Moreover, the qubits lose their saved state due to decoherence. 
Therefore, a quantum circuit does not generate correct results all the time when executed for a number of times. There are several metrics to quantify the \textit{correctness} of the output: (a) fidelity, (b) probability of basis/pure states, and (c) success rate. 
We briefly discuss these three metrics.

	\noindent\textbf{Fidelity:} Mathematically, fidelity between two quantum states, $\rho$ and $\sigma$ in density matrix representation, is expressed as \mbox{$F = \sqrt{\rho^{1/2} \sigma \rho^{1/2}}$}. This computes the closeness of the two density matrices. If the noise per operation is high, the actual output deviates more, and fidelity reduces. A higher fidelity is better in terms of reliability of a quantum circuit.

	\noindent\textbf{Probability of pure states:} In this approach, the output density matrix can be Eigen-decomposed with every possible pure states as the Eigen vectors. The Eigen values will then denote the probability of the pure state.

	\noindent\textbf{Probability of success:} This approach utilizes single-qubit and multi-qubit gate error-rates and law of probability to calculate the overall \textit{success rate} of a quantum circuit. Suppose, a quantum circuit consists of three gates, $G_1$, $G_2$, and $G_3$, each with error rate $\eta_1$, $\eta_2$, and $\eta_3$. According to the law of probability, the success rate of this circuit is $(1-\eta_1)(1-\eta_2)(1-\eta_3)$.

\begin{example}
Error rates for different qubits and qubit pairs of IBM QX 14 qubit quantum computer is shown in Fig.~\ref{tab:error}. It shows that multi-qubit gate error (CNOT) is an order higher than the single qubit gate error. Moreover, there is a qubit-to-qubit (Q2Q) variation among the error rates. For example, qubit pair $Q9-Q8$ has substantially higher error rate~$(0.32)$ than most of the other qubit pairs. Therefore, running operations on this qubit pair will be more erroneous compared to a qubit pair with lower error-rate, such as pair $Q1-Q0$.
\end{example}
In recent times, the concept of noise-aware mapping has emerged \cite{triq, tannu, ibm20q, qure, sabre}. There are qubit-to-qubit error-rate variations. The aforementioned works propose mapping and/or synthesis of a quantum circuit to less erroneous qubits for more reliable result. However, none of these works have considered all the constraints that we outline in this paper.

\subsection{Motivation}\label{ssec:back}
The recent works that consider the fidelity of quantum gates take a LNN circuit as a starting point~\cite{qure}. We present a motivational example in this subsection to demonstrate a range of design choices that remain unexplored, when decoupling these two steps of mapping.

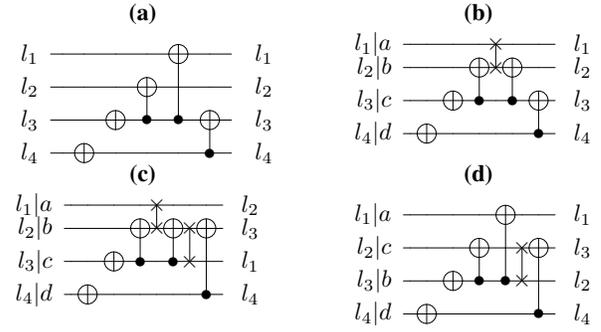
\begin{figure}[t]
	\centering
	\begin{subfigure}[t]{0.48\columnwidth}
        \centering 
	    \caption{}
	    \label{fig:orig_ckt}
		\scalebox{0.9}{
		\Qcircuit  @C= 0.5em @R=2mm{
			\lstick{l_1} & \qw & \qw & \qw & \qw & \targ & \qw & \qw &  	\rstick{l_1}\\
			\lstick{l_2} & \qw  & \qw & \qw & \targ & \qw & \qw & \qw & 	\rstick{l_2} \\
			\lstick{l_3} & \qw & \qw & \targ & \ctrl{-1} & \ctrl{-2} & \targ & \qw & 	\rstick{l_3}\\
			\lstick{l_4} & \qw & \targ & \qw & \qw & \qw & \ctrl{-1} & \qw & 	\rstick{l_4}
		}
	}
	\end{subfigure}
	\begin{subfigure}[t]{0.50\columnwidth}
		\centering
		\caption{}
		\label{fig:nn_l}
		\scalebox{0.9}{
			\Qcircuit @C= 1mm @R=2mm {
				\lstick{l_1|a} & \qw & \qw & \qw & \qw & \qswap  & \qw & \qw & \qw &  	\rstick{l_1}\\
				\lstick{l_2|b} & \qw  & \qw & \qw & \targ & \qswap \qwx &\targ & \qw & \qw & 	\rstick{l_2} \\
				\lstick{l_3|c} & \qw & \qw & \targ & \ctrl{-1} & \qw &\ctrl{-1} & \targ & \qw & 	\rstick{l_3}\\
				\lstick{l_4|d} & \qw & \targ & \qw & \qw & \qw & \qw & \ctrl{-1} & \qw & 	\rstick{l_4}
			}
		}
		
	\end{subfigure}
	
	\begin{subfigure}[t]{0.48\columnwidth}
		\centering
		\caption{}
		\label{fig:nn_t}
		\scalebox{0.9}{
			\Qcircuit  @C= 1mm @R=2mm {
				\lstick{l_1|a} & \qw & \qw & \qw & \qw & \qswap & \qw &  \qw  &  \qw & \qw &  	\rstick{l_2}\\
				\lstick{l_2|b} & \qw  & \qw & \qw & \targ & \qswap  \qwx & \targ & \qswap  & \targ & \qw & 	\rstick{l_3} \\
				\lstick{l_3|c} & \qw & \qw & \targ & \ctrl{-1} & \qw  & \ctrl{-1} &  \qswap \qwx & \qw  & \qw & 	\rstick{l_1}\\
				\lstick{l_4|d} & \qw & \targ & \qw & \qw & \qw & \qw &  \qw & \ctrl{-2} &  \qw & 	\rstick{l_4}
			}
		}
		
	\end{subfigure}
	\begin{subfigure}[t]{0.50\columnwidth}
		\centering
		\caption{}
		\label{fig:nn_grid}
		\scalebox{0.9}{
			\Qcircuit  @C= 1mm @R=2mm {
				\lstick{l_1|a} & \qw & \qw & \qw & \qw & \targ & \qw & \qw & \qw &  	\rstick{l_1}\\
				\lstick{l_2|c} & \qw  & \qw & \qw & \targ & \qw &  \qswap &\targ & \qw & 	\rstick{l_3} \\
				\lstick{l_3|b} & \qw & \qw & \targ & \ctrl{-1} & \ctrl{-2} &   \qswap \qwx & \qw& \qw & 	\rstick{l_2}\\
				\lstick{l_4|d} & \qw & \targ & \qw & \qw & \qw & \qw & \ctrl{-2} &  \qw & 	\rstick{l_4}
			}
		}
		
	\end{subfigure}
	\caption{(\subref{fig:orig_ckt}) Original Circuit. NN complaint  circuits considering  (\subref{fig:nn_l}) Linear topology (\subref{fig:nn_t}) T topology (\subref{fig:nn_grid}) Grid topology.}
	\label{fig:ckts}
\end{figure}
\begin{figure}[t]
	\begin{subfigure}[t]{0.4\columnwidth}
		\centering
		\caption{}
		\label{fig:nntopo}
		\includegraphics[width=2cm]{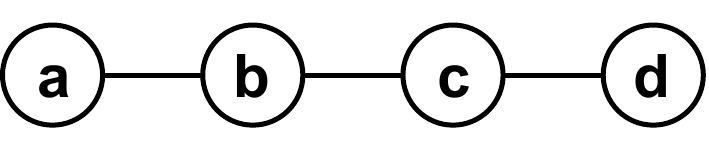}
	\end{subfigure}
\begin{subfigure}[t]{0.35\columnwidth}
		\centering
		\caption{}
		\label{fig:ttopo}
		\includegraphics[height=1cm]{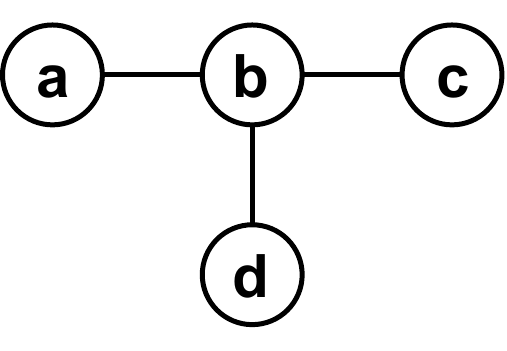}
	\end{subfigure}
	\begin{subfigure}[t]{0.22\columnwidth}
		\centering
		\caption{}
		\label{fig:gridtopo}
		\includegraphics[height=1cm]{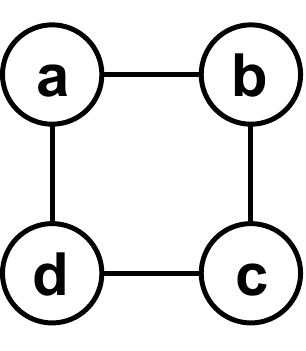}
	\end{subfigure}

\caption{3-qubit topology subgraphs extracted from topology graph presented in Fig.~\ref{fig:ibmq16}. (\subref{fig:nntopo}) Linear Topology
(\subref{fig:ttopo}) T topology
(\subref{fig:gridtopo})~Grid~topology.} 
\label{fig:topo}
\end{figure}	
Let us take the circuit presented in Fig.~\ref{fig:orig_ckt} as a representative example. The circuit comprises of four qubits~($l_1$, $l_2$, $l_3$, $l_4$). For this circuit to be mapped to a QC such as the one presented in Fig.~\ref{fig:ibmq16}, first we need to decide to which subgraph we want the circuit to be mapped to and make the circuit nearest neighbour compliant. 
\begin{enumerate}
	\item We consider 4-vertex non-isomorphic subgraphs extracted from the Fig.~\ref{fig:ibmq16}, as shown in Fig.~\ref{fig:topo}. 
	Each of these subgraphs can be used as the 'host' platform for the original circuit. The original circuit does not satisfy nearest neighbour constraint for any of the considered subgraphs and therefore, needs insertion of swap gates. Each topology~(subgraph) has different nearest neighbour constraints, which results in different NN-compliant circuits.  
	
	\item Given a subgraph $G$, we need to decide the starting configuration~$C$, i.e, mapping of a logical qubit~($l_1,l_2,l_3,l_4$) in the circuit to a unique vertex~($a,b,c,d$) in the subgraph. For example, Fig.~\ref{fig:nn_l} uses $l_1 \rightarrow a, l_2 \rightarrow b, l_3 \rightarrow c$ and $l_4 \rightarrow d$ as configuration.
	\begin{figure}[t]
    \centering
    \includegraphics[width=0.8\columnwidth]{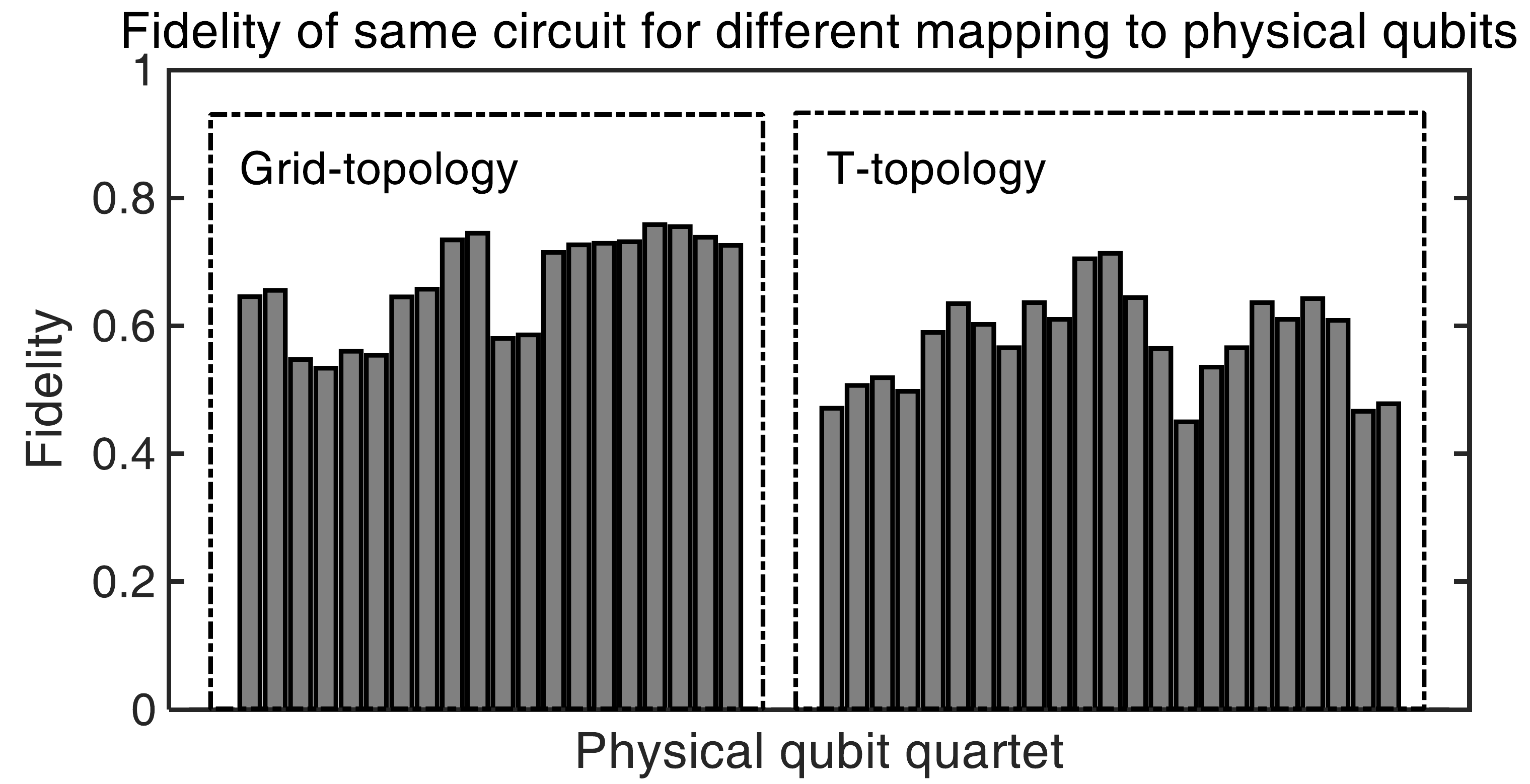}
    \caption{The same NN-compliant circuit mapped to different set of physical qubits result in different fidelity due to variations in qubit error-rates.}
    \label{fig:motivational}
    \vspace{-3mm}
\end{figure}

	\item With the chosen subgraph $G$ and initial configuration~$C$ of qubits, we construct the NN-circuit compliant with the chosen subgraph. Figure~\ref{fig:ckts} presents three different solutions, one for each topology subgraph. As evident from the figures, NN circuits of Fig.~\ref{fig:nn_l} and Fig.~\ref{fig:nn_grid} need a single SWAP gate for reaching NN compliance, while Fig.~\ref{fig:nn_t} requires two SWAP gates. 
	\item These NN-compliant circuits are subjected to further exploration of gate fidelity, produces different overall circuit resilience as can be observed in Fig.~\ref{fig:motivational}.
	
	As evident from Fig.~\ref{tab:error}, qubits have different error-rates. Instead of scheduling the NN compliant circuit in an error-agnostic fashion to any set of qubits, qubits with better error-rates can be selected to execute the circuit. For example, the qubits \{a, b, c, d\} in the T-topology in Fig. \ref{fig:nn_t} can be assigned to a number of different sets physical qubits (suppose, \{0, 1, 2, 13\} and \{8, 9, 10, 5\}) in the 14-qubit IBM computer (Fig. \ref{fig:ibmq16}). Although both assignment will satisfy NN compliance, they are not equivalent in terms of operation fidelity. Gate operations involving qubit pair $Q9-Q8$ and $Q9-Q10$ introduces a higher error compared operations on pair $Q1-Q0$ and $Q1-Q2$ due to different gate error-rates (Table \ref{tab:error}). Therefore, in this step error-rate difference of qubits is taken into consideration, and the NN compliant circuit is mapped to better quality qubits for a better noise resiliency.
\end{enumerate}

\section{Methodology}\label{sec:method}
\begin{figure}
    \centering
    \includegraphics[width=6cm]{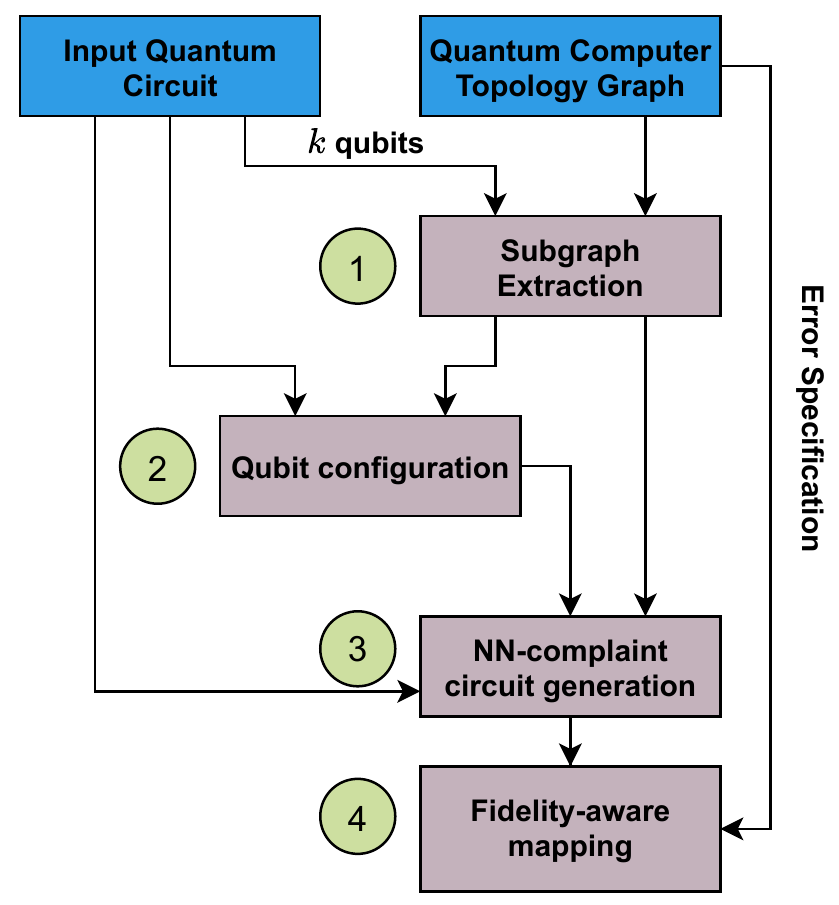}
    \caption{Proposed Multi-constraint Quantum Circuit Mapping to NISQ computer.}
    \label{fig:techflow}
\end{figure}
\noindent From the demonstrative example, we observe the following degrees of freedom, applicable to the technology mapping, when starting from a quantum circuit without any neighbourhood compliance.

\begin{enumerate}[leftmargin=*]
	\item \emph{Topology Subgraph Selection}: A quantum computer with $n$ qubits can host a quantum circuit with set of gates operating on $k$ qubits, such that $k \leq n$ using different subgraphs in it, as depicted in the Fig.~\ref{fig:topo}.
	\item \emph{Logical Qubit to Topology graph node Mapping}: For a given subgraph extracted from the `host' quantum computer,  there remains the possibility of different qubit to subgraph vertex mappings. We define this as a \emph{configuration}, presented in Definition~\ref{def:config}.
	\item \emph{Nearest Neighbour~(NN) Compliance}: Given an initial configuration and topology subgraph, NN compliance can be achieved by inserting swap gates, with the objectives to minimize delay as well number of swap gates required for NN compliance.
	\item \emph{Fidelity-aware mapping of NN-compliant circuit to QC}:
	We determine the mapping of the  NN-compliant circuit to the qubits in the quantum computer, while considering the qubit and gate error rates to minimize error-rate.
\end{enumerate}

The above degrees of freedom can be independently or jointly exercised to optimize one or more of the following targets - gate count, logical depth and circuit fidelity. In this work, we focus on achieving high circuit fidelity. The proposed technology mapping flow is shown in Fig.~\ref{fig:techflow}. We describe the individual blocks in the following subsections. 

\subsection{Extracting Topology Graph}
Given a quantum circuit over $k$-qubits and a NISQ computer with $n$~qubit, such that $n>k$, there are multiple possible embedding of the circuit onto the NISQ computer. We use a probabilistic algorithm with fixed number of \emph{attempts} to extract topology sub-graphs from a given quantum computer topology graph~$T$, as shown in Algorithm~\ref{algo:subgraph}.

\begin{algorithm}
\SetKwData{Left}{left}
\SetKwData{This}{this}
\SetKwData{Up}{up}
\SetKwFunction{Union}{Union}
\SetKwFunction{List}{List}
\SetKwFunction{iso}{IsIsomorphic}
\SetKwRepeat{Do}{do}{while}
\SetKwFunction{FindCompress}{FindCompress}
\SetKwInOut{Input}{input}
\SetKwInOut{Output}{output}
\Input{Topology Graph $T$, $k$, $attempts$}
\Output{$L$ :A list of $k$-vertex non-isomorphic subgraphs.}
\BlankLine
L = \List{}\;
\For{$i\leftarrow 1$ \KwTo $attempts$}{
\tcc{Generate a subgraph with $k$ vertices}
$v$ = Pick a vertex from $T$ randomly\;
$N.add(v)$\;
    \While{\texttt{not} $N.empty()$}{
        $v_N = N.pop()$\;
        \If{$v_N \notin |g_{new}.V|$}{ 
            $g_{new}.V.add(v_N)$\;
            $N_{new}$ = Choose neighbours of $v \in T$ with probability $p$, not considered before\;
            $N = N+N_{new}$\;
        }
        \If{$|g_{new}.V| == k$}{break\;}
        
     }
    Add edges from $T$ to $g_{new}$ for the subgraph induced by $g_{new}.V$\;
    
\If{\texttt{not} \iso{$g_{new}, L$}}{L.add($g_{new}$)}
}
\caption{$k$-vertex Topology graph extraction}\label{algo:subgraph}
\end{algorithm}

\subsection{NN-compliant circuit technology mapping}
In this subsection, we present an optimal technique based on Integer Linear Programming~(ILP) formulation for solving the NN-compliant circuit mapping problem for arbitrary topologies~\cite{bhattacharjee2017depth}. We consider an input circuit over say $n$~qubits to be mapped with a topology graph with $n$~vertices extracted using the technique presented in the previous subsection. We then generate some random configurations mapping each qubit to a vertex and solve the NN-compliant technology mapping problem. Formally, a configuration can be defined as follows. 
\begin{definition}(Configuration)\label{def:config}
	A \textbf{configuration} $C_t$ is the set of ordered tuples $(q_i, v)$, which indicates that in cycle $t$, qubit $q_i$,  is at location $v$, $ 1\le i \le n$ and $v \in T_\mathcal{V}$.
	Configuration $C_0$ represents the initial configuration.
\end{definition}

 Given an initial configuration $C$ of $n$-inputs, a series of levels $L_1, L_2, \ldots, L_k$ and a topology graph $T$, the objective is to determine the series of swap gates needed to transform the location of the qubits from configuration $C$ such that all qubits pairs in interaction $I_1$~(corresponding on level $L_1$) are nearest-neighbor, and then again location of qubits are transformed to be nearest neighbors for $I_2$~(corresponding on level $L_2$) and so on, till interaction $I_k$~(corresponding on level $L_k$) is met and the combined delay of swap gates and gates present in the actual circuit is minimum. 

\begin{table}
\centering
\caption{Parameters/constants used in ILP}
\label{table:param}
{\small
\begin{tabular}{cl}\bottomrule
\textbf{Param/const.} & \textbf{Description}  \\ \midrule
$G$  & Toplogy graph \\ 
$C$ & Input/start configuration \\ 
$n$ & Number of inputs \\ 
$k+1$ & Number of levels \\ 
$L_i$ & Number of qubit interaction pairs in level $i$ \\
$T$ & Maximum number of cycles used for the problem \\ \toprule
\end{tabular}
}
\end{table}


\begin{table}
\centering
\caption{Variable description used in ILP}
\label{table:var}
{\small
\begin{tabular}{|c|p{6cm}|}\hline
\textbf{Var.} & \textbf{Description}  \\\hline
$delay$ & Delay due to insertion of swap gates \\ \hline$c_{v,q,t}$  & 1 indicates qubit $q$ will move to new location $v$ in cycle $t$ \\ \hline
$m_{i,t}$ & 0 indicates Interaction $i$ met in cycle $t$ \\\hline$a_{i,t}$ & 1 indicates gates in Level $i$ are scheduled in cycle $t$ \\ \hline
$n_{p,q,t}$ & 1 indicates qubit $p$ and $q$ are NN in cycle $t$ \\\hline$eb_{I_i,t}$ & 1 indicates interaction $I_i$ has been met in cycle $t$ and gates of level $i$ can be placed in the current or following cycles. \\ \hline
$p_{(p,v),(q,w),t}$ & 1 indicates qubit $p$ is in location $v$ and $q$ is in location $w$ in cycle $t$ \\\hline$b_{q,t}$ & 1 indicates qubit $q$ cannot be involved in a swap in cycle $t$ \\ \hline
$x_{v,p,t}$ & 1 indicates qubit $p$ is in location $v$ in cycle $t$ \\\hline$b_{v,q,t}$ & 1 indicates qubit $q$ in location $v$ cannot be involved in a swap in cycle $t$ \\ \hline
$u_{v,q,t}$ & 1 indicates qubit $q$ will remain in location $v$ in cycle $t$ \\\hline$sb_{m,n,t}$ & 1 indicates swap is not permitted between locations $m$ and $n$ in cycle $t$ \\ \hline
\end{tabular}
}
\vspace{-3mm}
\end{table}

\begin{figure*}
\includegraphics[width=\textwidth]{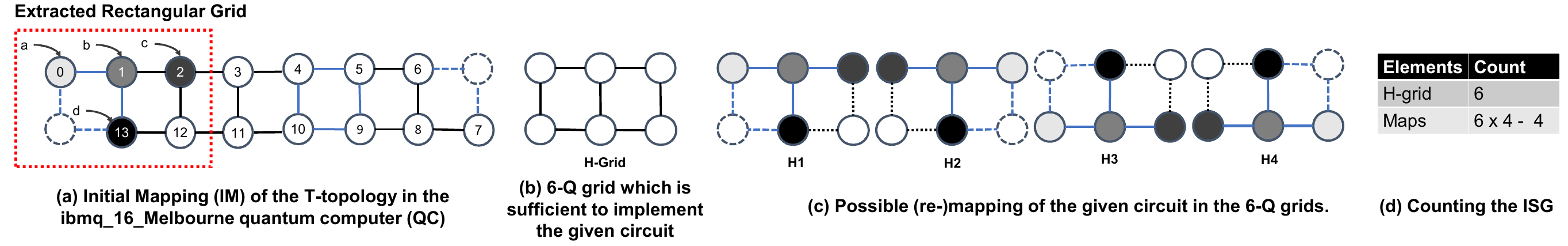}
\caption{Methodology for finding isomorphic sub-graphs of the NN-compliant circuit to compute fidelities.}
\label{fig:qure}
\vspace{-5mm}
\end{figure*}
\head{Objective function:}
\begin{equation}\text{Minimize }\sum_{i=0}^k\sum_{t = 0}^T t. a_{i,t} \end{equation}

\head{Level scheduling constraints:} Each level can be scheduled/activated exactly once.
\begin{align}
 \sum_{t=0}^{T} a_{i,t} &= 1; &0 \le i \le k
\end{align}
\noindent Only one level can be activated per time step.
\begin{align}
 \sum_{i=0}^{k} a_{i,t} &= 1; &0 \le t \le T
\end{align}
\noindent Activation for a level $i$ can happen only if corresponding interaction $i$ is met.
\begin{align}
 a_{i,t} + m_{i,t} &\le 1; &0 \le t \le T, 0 \le i \le k
\end{align}

\head{Swap blocking constraints:}
If an interaction $i'$ is met and all the gates in any Level $i$ such that $(i < i')$ have been scheduled, then swaps involving the qubits in interaction $i$ cannot be performed and interaction $i'$
is blocked till Level $i$ has been scheduled. Qubit involved in an interaction $i$ cannot be swapped in the cycle, when the Level $i$ is scheduled.
\begin{align}
eb_{i',t} &= a_{i,t} \wedge (1 - m_{i',t}) \nonumber \\
0 &\le i \le k-1, i+1 \le i' \le k, 0 \le t \le T \\ 
b_{q,t} &= \vee_{i} (a_{i,t} \vee eb_{i,t}) ~~~\forall i~\exists ~q \in I_i, 0 \le t \le T \\
 b_{v,q,t} &= b_{q,t} \wedge x_{v,q,t} ~~~0 \le t \le T \\
 sb_{m,n,t} &= \vee_q (b_{m,q,t} \vee b_{n,q,t}) ~~~\forall q \in Q , 0 \le t \le T 
\end{align}

\head{Chronological interaction constraints:} If an interaction is met in cycle $t$, then the status should not change to not met after
that cycle. In addition, interaction $i$ must be met before ${i-1}^{th}$ interaction is met.
\begin{align}
m_{i,t+1} - m_{i,t} &\ge 0 & 0 \le t \le T-1, 0 \le i \le k \\  
m_{i+1,t} -  m_{i,t} &\ge 0 &0 \le t \le T, 0 \le i \le k-1
\end{align}

\head{Successful interaction constraints:} An interaction is met if all the qubit pairs in the interaction are nearest neighbors. If an interaction has been met in
cycle $t$, then in all cycles $t' > t$, the qubit positions do not matter any longer.
\begin{align}
L_i.m_{i,t} + (\sum_{(p,q) \in I_i} n_{p,q,t}) + (\sum_{t'=0}^{t-1}L_i.(1 - m_{i,t'})) &\ge L_i \nonumber \\ 0\le t \le T
\end{align}

\head{Nearest neighbor constraints:} Two qubits $p$ and $q$ are nearest neighbors if the qubits are in two locations $v$ and $w$ respectively or in $w$ and $v$ respectively, such that $(v,w) \in G_{\mathcal{E}}$ and $(p,q) \in I$.
\begin{align}
p_{(p,v),(q,v),t} &= x_{v,p,t} \wedge x_{w,q,t}  \\
p_{(p,w),(q,v),t} &= x_{w,p,t} \wedge x_{v,q,t}; \\ 
n_{p,q,t} &= \vee_{(v,w) \in G_{\mathcal{E}}} (p_{(p,v),(q,w),t} \vee p_{(p,w),(q,v),t})
\end{align}

\head{Qubit position update constraints:} A qubit $q$ is at location $v$ in cycle $t+1$ if it was in location $v$ in cycle $t$ and there were no swaps performed involving the location $v$ or
if $q$ was in a location $w$ which is nearest neighbor with $v$ and a swap was performed between $v$ and $w$.
\begin{align}
 u_{v,q,t+1} &= (\wedge_{(v,w) \in G_{\mathcal{E}}} (1 - s_{v,w,t}))\wedge x_{v,q,t};  \\
 c_{v,q,t+1} &= \vee_{(v,w) \in G_{\mathcal{E}}} s_{v,w,t}~\wedge ~x_{w,q,t}\\
 x_{v,q,t+1} &= u_{v,q,t+1} ~\vee~ c_{v,q,t+1}
\end{align}

\head{Qubit location and swap constraints:} A qubit $q$ can be at exactly one position in any given cycle. In a given cycle, a location can be involved in at most one swap.
\begin{align}
 \sum_{v \in G_{\mathcal{V}}} x_{v,q,t} &= 1; &0 \le t \le T, q \in Q \\
 \sum_{(v,w) \in G_{\mathcal{E}}} s_{v,w,t} &\le 1; &0 \le t \le T, v \in G_{\mathcal{V}}
\end{align} 
\head{Initialization constraints:} A qubit~$q$ is at location $v$ in cycle 0, based on input configuration $C$.
\begin{align}
 x_{v,q,0} &= 1; &(v,q) \in C
\end{align}
This completes the description of the ILP formulation of NN-mapping of quantum circuits for arbitrary topologies. The variables used in the ILP formulation are summarily presented in Table~\ref{table:var}. For large circuits, the ILP solves takes a long time to find the optimal solutions. Therefore, we split the input circuit into non-overlapping windows. A window can be defined as a set of consecutive levels in a circuit. For example, a window size of $2$ would split a circuit with $l$ levels into $\frac{n}{2}$ windows. We solve the NN compliance for each window separately. 

\subsection{Fidelity-aware mapping}
The NN-compliant circuit generated from the previous step is further optimized by taking qubit error-rate variations into account. As demonstrated in the motivational example, the same NN-compliant circuit when mapped to a different set of physical qubits on quantum computer can result in different fidelities or probability of successes. To find a mapping of the NN-compliant circuit to less erroneous physical qubits, we exploit the regularities in existing NISQ devices. The NISQ devices generally follow a grid-like architecture (e.g., IBMQ16)\footnote{Note that the topology-aware mapping formulation through ILP does not restrict the host quantum architecture to have a grid-like formation}. 

\noindent \emph{Step 1.} After getting the NN-compliant circuit, we extract a rectangular grid architecture from the coupling graph which is sufficient to implement the given quantum circuit. As an example, we take the T-topology in  Fig. \ref{fig:qure}(a) as the NN-circuit and IBMQ16 as the target NISQ computer. We call this H-Grid. 
Inside the H-Grid, the given workload can be mapped in at least 4 different ways as shown in Fig. \ref{fig:qure}(c). These are basically the horizontally and vertically mirrored re-assignment of the qubits from initial rectangular grid within the H-Grid.
\begin{figure*}[ht]
\centering
\begin{minipage}{\textwidth}
\centering
\includegraphics[width=0.90\textwidth]{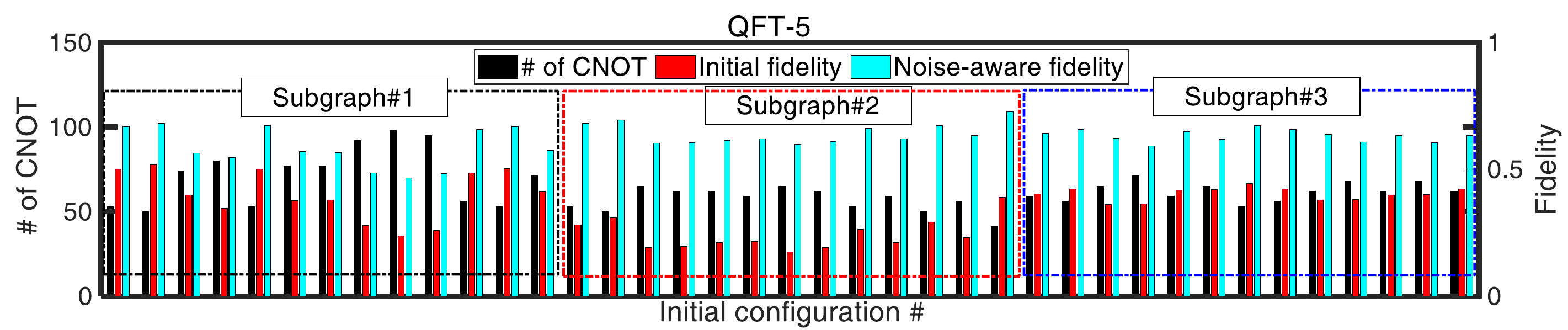}
\caption{The number of CNOT gates and the fidelity of the QFT-5 circuit show variation with respect to initial topology graph and qubit configuration. One choice of initial topology can generate a smaller number of gates and a higher fidelity than others. Moreover, the fidelity of the initial placement can be further improved through fidelity-aware mapping (avg. $1.92$x better).}
\label{fig:qft5-all}
\end{minipage}

\vspace{-5mm}
\end{figure*}[t]



\begin{figure*}[t]
\begin{subfigure}{0.32\textwidth}
\centering
\includegraphics[width=\textwidth]{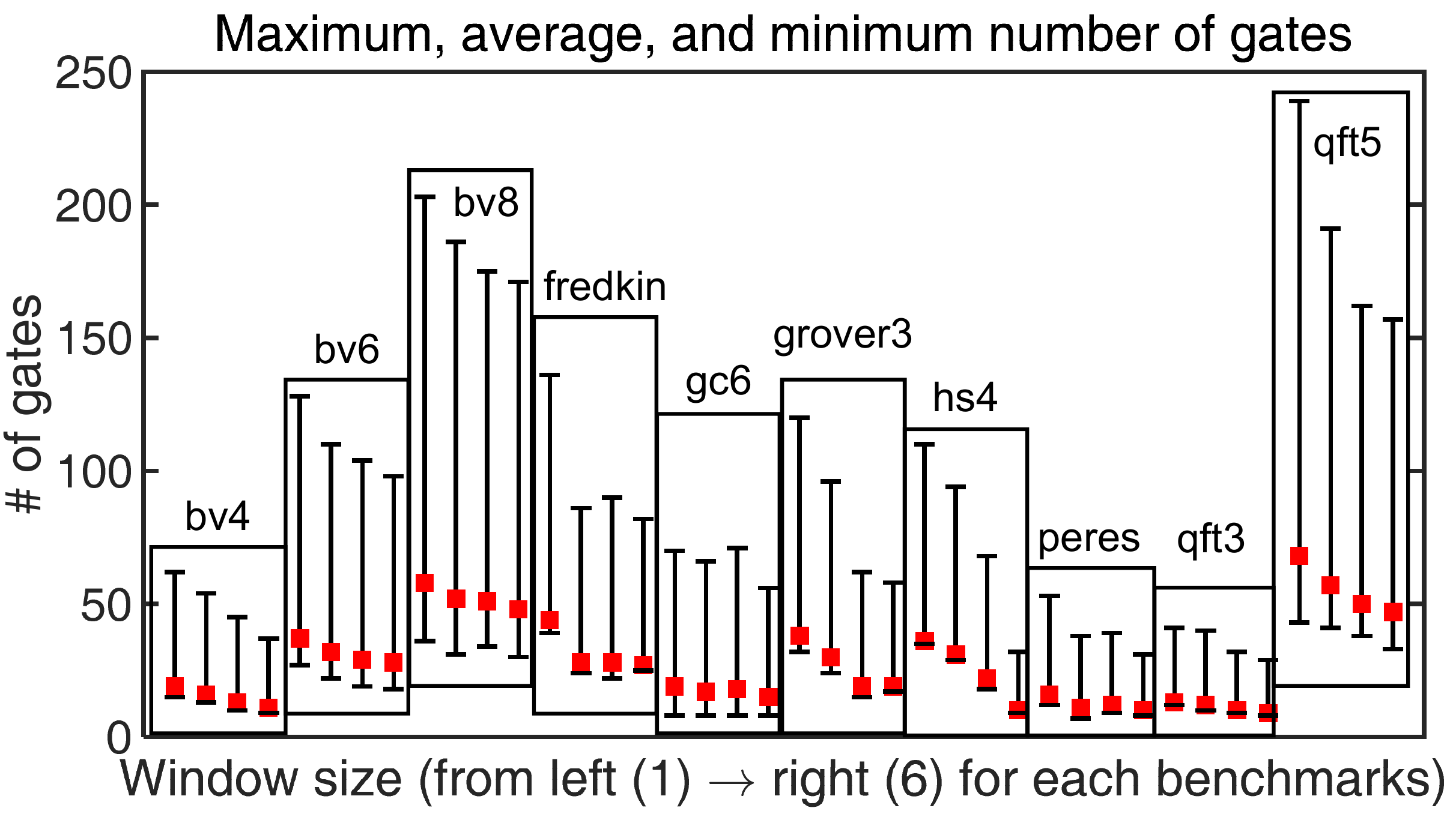}
\caption{}
\label{fig:global-gate-count}
\end{subfigure}
\begin{subfigure}{0.32\textwidth}
\centering
\includegraphics[width=\textwidth]{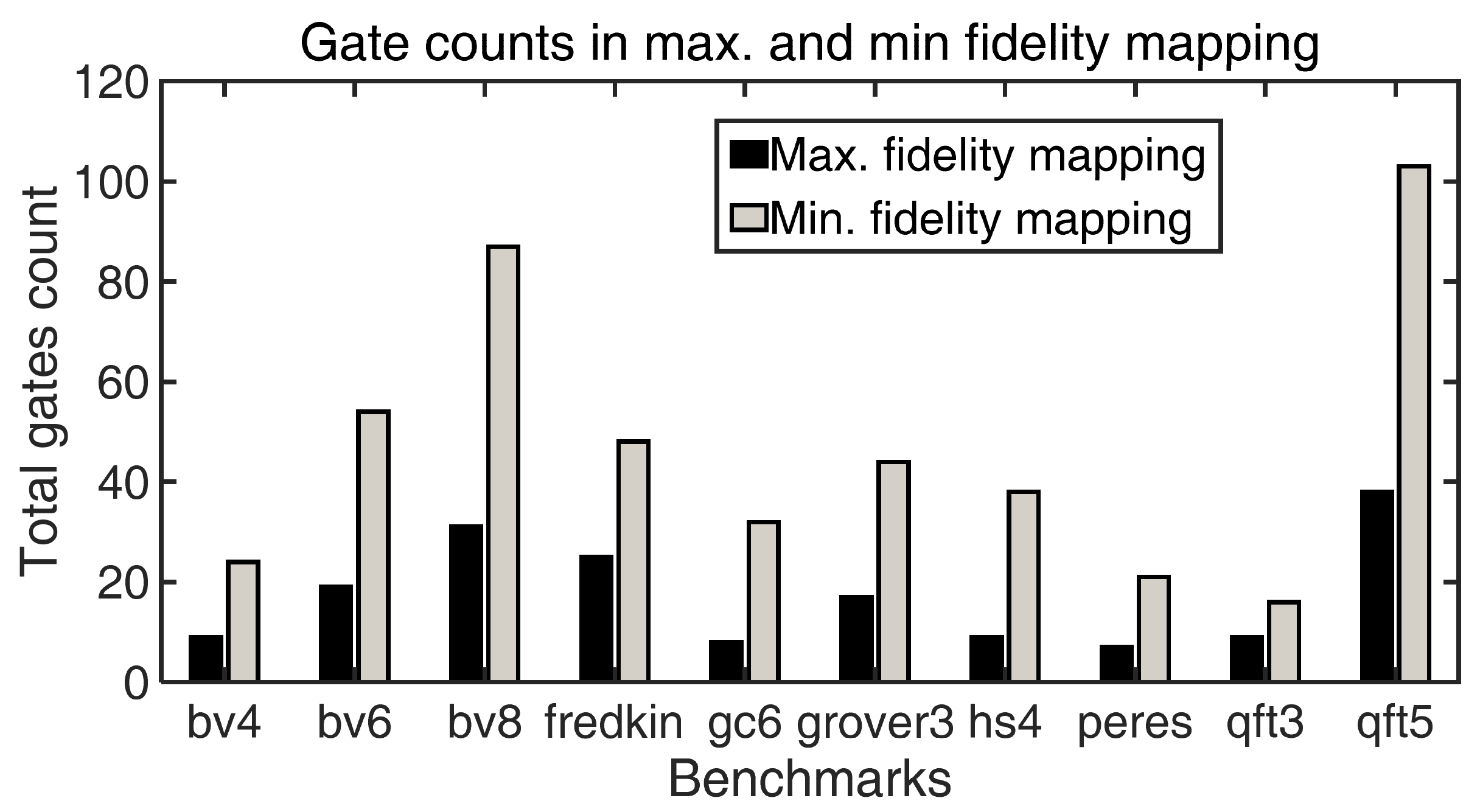}
\caption{}
\label{fig:global-gate-count}
\end{subfigure}
\begin{subfigure}{0.32\textwidth}
\centering
\includegraphics[width=\textwidth]{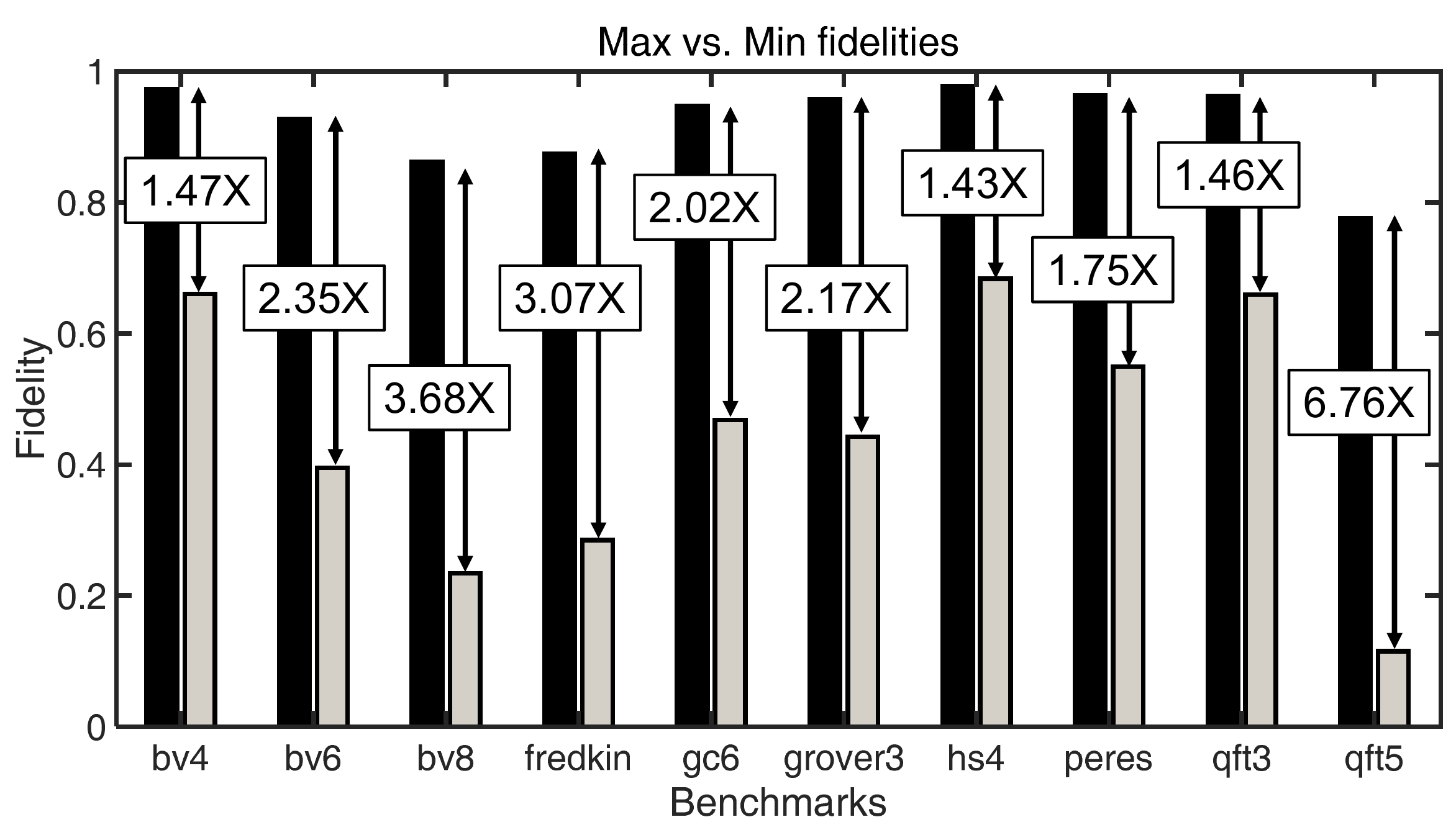}
\caption{}
\label{fig:global-fidelity}
\end{subfigure}
\caption{(a) Selection of window size in the NN topology solver has an effect on the number of gates in the final circuit. In general, a larger window size  $\equiv$ a smaller \# of gates, (b) max. fidelity mappings have a smaller \# of gates than min. fidelity maps and (b) Improvement of fidelities across the degree of freedoms. By exploring topology graph, qubit configurations, and noise-awareness, the final fidelity of a quantum circuit can be improved substantially.}
\vspace{-6mm}
\end{figure*}

\begin{figure*}
\centering
\begin{subfigure}{0.32\textwidth}
\centering
\includegraphics[width=\textwidth]{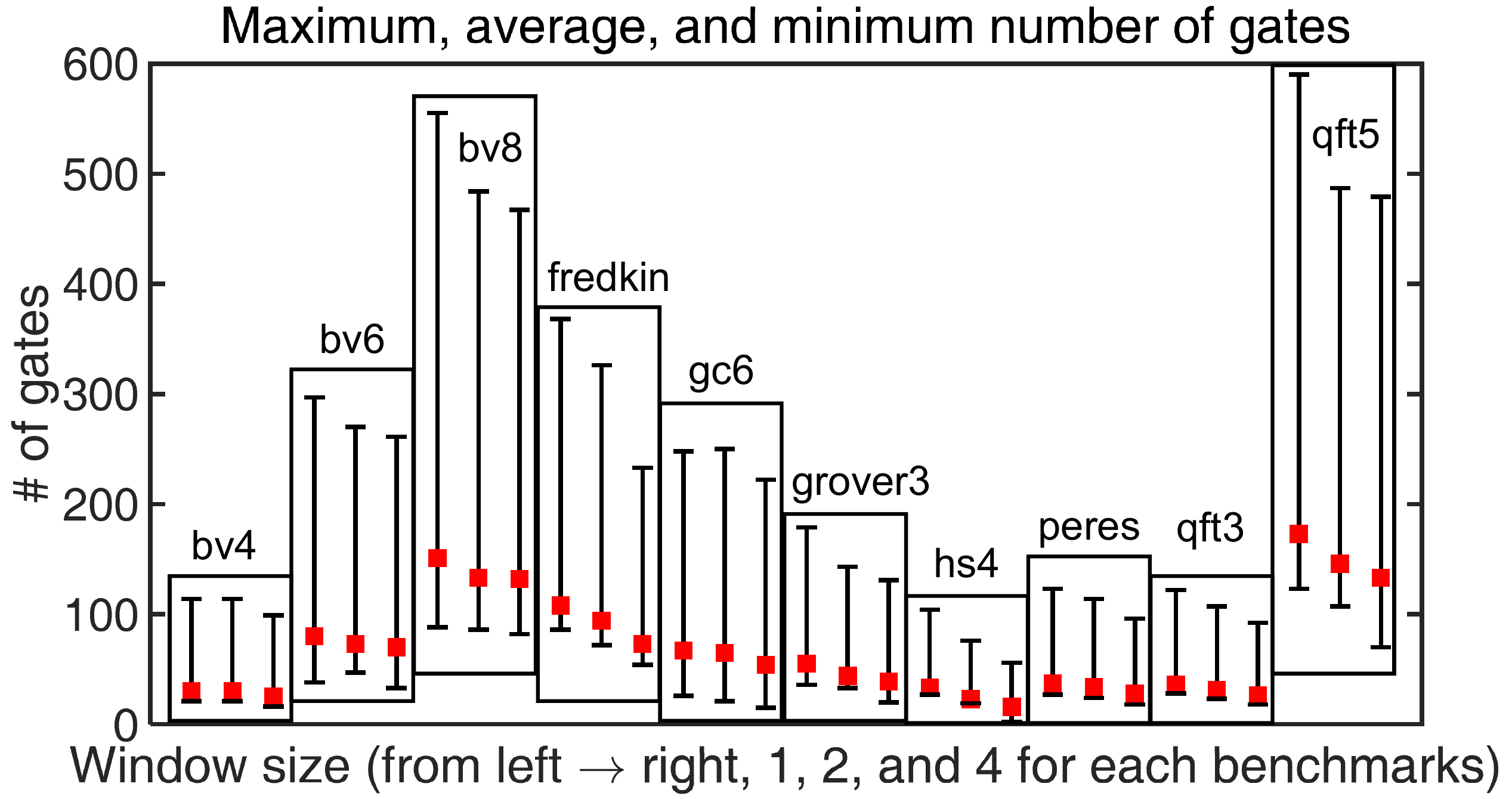}
\caption{}
\label{fig:rigetti-errorbar}
\end{subfigure}
\begin{subfigure}{0.32\textwidth}
\centering
\includegraphics[width=\textwidth]{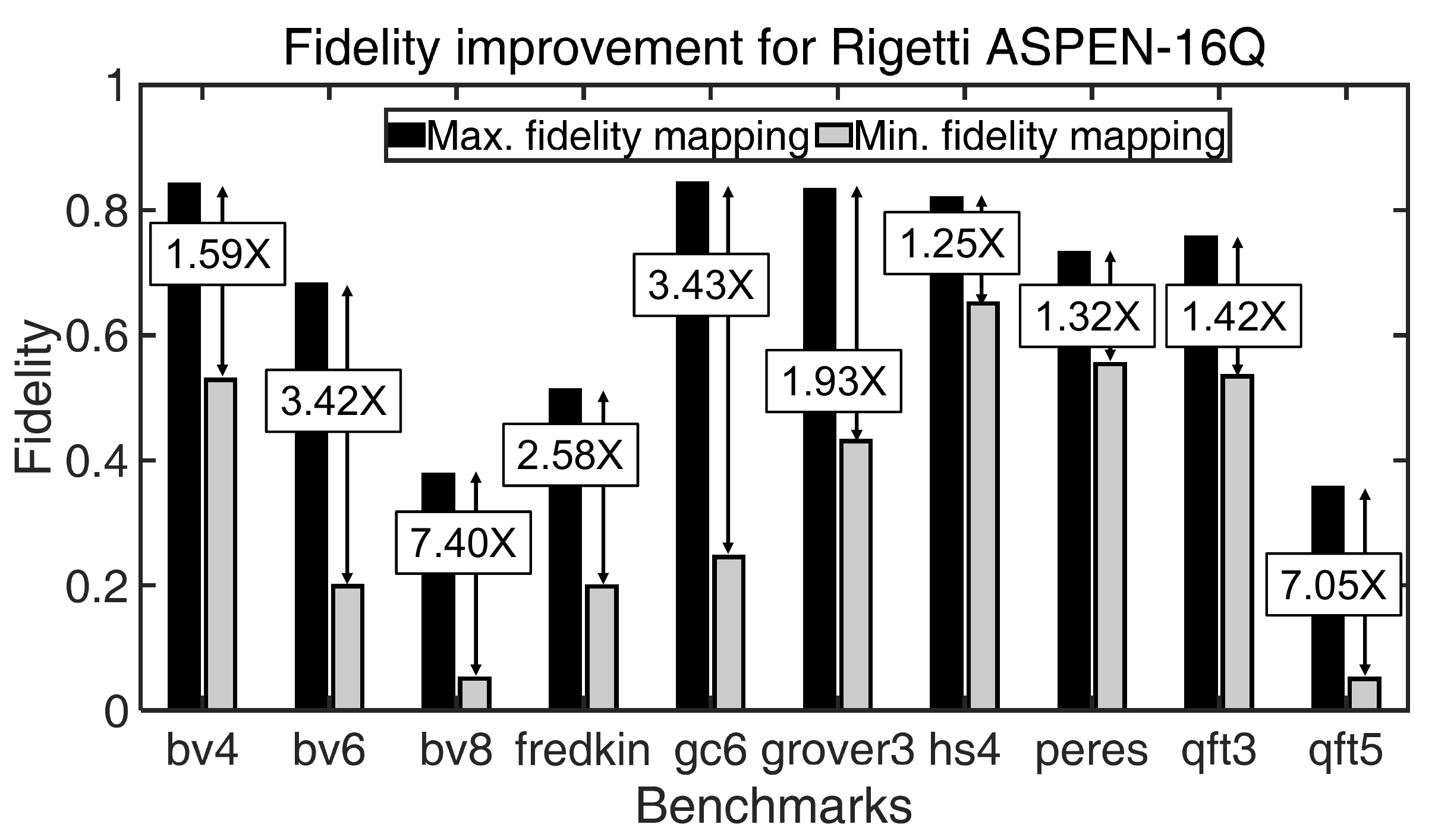}
\caption{}
\label{fig:rigetti-fidelity}
\end{subfigure}
\caption{Results of mapping for Rigetti ASPEN 16Q computer. (b) Maximum, average, and minimum number of gates in the NN-compliant circuit. Typically, a larger window size results in a smaller \# of gates, and (c) improvement of fidelity. Results show depending on the selection of window and initial topology graph, the fidelity can vary substantially.}
\label{fig:rigetti}
\vspace{-6mm}
\end{figure*}
\noindent \emph{Step 2.} We determine the uniquely fitted H-Grids within the $QC$ by sliding the rectangular grid over the $QC$. Let the number of qubits in the horizontal and vertical direction are $QH$, and $QV$ for the $QC$, and $HQH$ and $HQV$ for the H-Grid. Then the number of H-Grid ($NHG$) within the $QC$ will be (QH-HQH+1)*(QV-HQV+1).

\noindent \emph{Step 3.} We compute all possible mapping coordinates for the qubits within each unique H-Grid. The total number of unique mapping ($NISG$) in the target hardware for the given workload will be at least ($4 \times NHG$) for $HQH$, $HQV$ $>$ 1. Finally, for each possible mapping fidelity is computed taking the qubit error-rates into consideration. 


\section{Experimental Results}\label{sec:res}
The proposed multi-constrained technology mapping flow\footnote{https://github.com/debjyoti0891/quantum-chain} was implemented in Python, with Gurobi~8 as the ILP solver~\cite{gurobi}. 



\noindent \textbf{Topology graph variation:} We demonstrate the effect of topology graph variation with QFT5 benchmark. Fig. \ref{fig:qft5-all} shows the number of CNOT gates in the final NN-compliant circuit and the fidelity of the initial random mapping for different topology graphs and qubit configurations (window size~$w=1$). For space constraint, we show 39 out of 78 configurations for QFT-5 benchmark. However, for clarity we have plotted 39 of those). Results show that depending on the qubit configuration and topology graph, the final circuit can have different number of CNOT gates to satisfy the NN compliance. A smaller number of CNOT gates or the total number of gates in general, results in a better fidelity. 

However, the initial placement of the NN-compliant circuit 
does not take the qubit-to-qubit error-rate variations in account. Therefore, the fidelity-aware mapping step further searches for a mapping with better fidelity. In Fig.~\ref{fig:qft5-all}, the \textit{noise-aware fidelity} bar shows that  noise-awareness in conjunction with gate-reduction can improve the fidelity. We observe on average $1.96\times$ improvement for the QFT-5 benchmark. 

\noindent\textbf{Variants of NN topology solver:} We explore the technology mapping flow with various window sizes in the NN topology solver,  $w$ = \{$1$, $2$, $4$, $6$\}. The maximum, average, and minimum number of total gates are plotted for different benchmarks for these window sizes. The total number of gates consists of noisy gates only from the IBM QX quantum computer, i.e., U2, U3, and CNOT gate. With larger window sizes, the NN topology solver can reduce the overall number of gates required due to optimization across multiple levels in the circuit, to find a NN complaint circuit.

\noindent\textbf{Putting it all together:} The proposed flow offers the scope of optimizing multiple parameters. Providing a large window size as input to the topology solver results in a NN-compliant circuit with fewer gates, at the cost of higher run times. For all the benchmarks, the mapped circuit with maximum fidelity has a substantially smaller number of gates compared to the minimum fidelity mapping of the same circuit, as evident from Fig.~\ref{fig:global-gate-count}. This is intuitively correct since the fidelity of a circuit is inversely related to the number of gates in the circuit.  The simulation results with realistic error-rates and connectivity of IBMQ16 shows that the overall fidelity can be improved up to $6.76\times$ among the simulated benchmarks, as shown in Fig.~\ref{fig:global-fidelity}. The most improvement is observed for QFT$5$ benchmarks which has the largest number of gates among the simulated benchmarks. Therefore, our proposed flow offers a myriad of design space choices~(window-size, initial topology, qubit configuatioo, and noise-aware mapping to physical qubits) to improve the fidelity of a given input quantum circuit. 

\noindent\textbf{Variants of quantum computer:} The proposed flow also permits mapping to various quantum computers. To demonstratet this, we consider the of Rigetti's 16-qubit quantum computer that has a topology graph as shown in Fig.~\ref{fig:rig}. Note that, most of the qubits are connected to 2 other qubits. The native 2-qubit gate of Rigetti system is Controlled-Z (CZ). Decomposing SWAP gate with native Rigetti gates results in quantum circuit with 18 gates of which 11 are noisy RX and CZ, and 7 are noise-free RZ. Due to large number native operations for a SWAP gate, the total number of gates in the NN-compliant circuits are relatively larger for Rigetti system.  We generate NN-compliant circuit for the same benchmarks and plot the maximum, average, and minimum number of gates for each benchmark for different window sizes ($w=\{1, 2, 4\}$). Fig.~\ref{fig:rigetti-fidelity} shows the maximum and minimum fidelity of the NN-compliant circuits for each benchmark. Rigetti reports mean 1-qubit (F1Q) and 2-qubit (F2Q) gate fidelities instead of qubit-wise specifications. Therefore, for the analysis in Fig. \ref{fig:rigetti-fidelity} all the qubits are considered identical with same gate fidelities (Accessed: 06-Aug-2019; F1Q = 93.79\% and F2Q = 91.71\%). However, in reality, gate error-rates will exhibit qubit-to-qubit variation, which offer the scope for additional improvement in fidelity over the reported results in Fig.~\ref{fig:rigetti-fidelity}.




\section{Conclusion}\label{sec:conc}
In this paper, we presented an integrated flow for multi-constraint quantum circuit mapping on noisy intermediate-scale quantum computers. The flow explores a number of degrees of freedom including various topology graphs, qubit mapping, NN compliance, and qubit error-rates to generate mapping with a high fidelity. An optimal integrated mapping remains to be explored in future, along with studies for larger benchmarks.

\textbf{Acknowledgement: }
{\scriptsize SRC task 2847.001,  NSF (CNS- 1722557, CCF-1718474, DGE-1723687 and DGE-1821766) and DARPA Young Faculty Award (D15AP00089).}
\bibliographystyle{unsrt}
\bibliography{ref}
\end{document}